\documentclass[10.5pt]{iopart}

\usepackage{graphicx}
\usepackage{bm}
\usepackage{appendix}
\usepackage{color}
\usepackage{braket}

\usepackage{multicol}
\usepackage{amsmath}
\usepackage{cite}

\begin{document}

\title[]{
Pseudo anomalous Hall effect in semiconductors and semimetals: A classical perspective
}

\author{Akiyoshi Yamada}
\author{Yuki Fuseya}%

\address{Department of Physics, Kobe University, Kobe 657-8501, Japan\\}
\ead{a\_yamada@people.kobe-u.ac.jp}
\vspace{10pt}
\begin{indented}
\item[]\date{\today}
\end{indented}

\begin{abstract}
We demonstrate that the non-linear field dependence in the Hall effect, often indistinguishable from the anomalous Hall effect, can be realized entirely within the classical mechanism due to the Lorentz force by analyzing multi-valley models for semiconductors and semimetals.
The non-linear component in the Hall resistivity $\rho_H^{\rm NL}$ originates from carrier mobility anisotropy or the coexistence of different charges. Since $\rho_H^{\rm NL}$ is inversely proportional to the carrier difference between electrons and holes $\Delta n$, it exceeds its zero-field value near charge neutrality.
As a practical example, we show that the magnitude of the classical non-linear Hall response in ZrTe$_5$ is comparable to the experimental values, underscoring the importance of accounting for classical contributions before attributing non-linear Hall effects to quantum mechanisms.
\end{abstract}

\maketitle
\ioptwocol

\section{Introduction}

The anomalous Hall effect (AHE) is a fascinating consequence of the adiabatic phase (the Berry phase~\cite{Berry1984}) acquired by the wave function~\cite{KL1954,BERGER19641141,SMIT1955877,SMIT195839,nagaosa2010}. 
The AHE arises under the fundamental condition of broken time-reversal symmetry in the presence of spin-orbit coupling~\cite{KL1954,Crepieux2001,nagaosa2010}. 
While the AHE has been extensively studied in ferromagnetic metals~\cite{nagaosa2010}, its recent observation in antiferromagnetic metals~\cite{vsmejkal2022anomalous} has revitalized interest in this field, opening new avenues for exploring Hall effects in solid state physics.

Experimentally, the AHE is typically measured under an external magnetic field, where the normal Hall effect, driven by the Lorentz force, inevitably coexists. 
The conventional approach for isolating the AHE involves subtracting the normal Hall component, assuming a monotonic linear dependence on the magnetic field. 
However, this method introduces intrinsic ambiguity: if the normal Hall effect exhibits non-linear field dependence --- a behavior commonly observed in semiconductors and semimetals, and we call it "pseudo anomalous Hall effect (PAHE)" --- this procedure can lead to erroneous conclusions. 
Despite its experimental significance, the precise mechanisms underlying non-linear field dependence in the normal Hall effect remain poorly understood even today.

In this paper, we clarify the origins and characteristics of non-linear field dependence in the normal Hall effect. Specifically, we address the following fundamental questions, which have so far lacked precise answers:
(i) Under what conditions does the Lorentz force induce non-linear field dependence?
(ii) How significant are these non-linear components in typical experiments?
(iii) How can we isolate and eliminate these components experimentally?
In the following sections, we provide definitive answers to these questions by analyzing three representative systems: (A) anisotropic multivalley semiconductors, (B) semimetals, and (C) ZrTe$_5$ as an example of an anisotropic semimetal. Our findings shed new light on the fundamental nature of the non-linear normal Hall effect and its implications for experimental studies of the AHE. 

\section{Non-linear component by multi-valley model}\label{sec_Hall}
The contributions to the magnetoconductivity from the normal Hall effect are evaluated using classical transport theory --- the Drude theory. 
The magneto-conductivity tensor in an ellipsoidal Fermi surface is expressed as~\cite{MacKey1969,Aubrey1971,Zhu2018,Mitani_2020}
\begin{eqnarray}
\hat{\sigma}_i&=&e n_i\left(\hat{\mu_i}^{-1}\pm\hat{B} \right)^{-1},\label{eq_conductivity}
\end{eqnarray}
in terms of the $3\times 3$ magnetic field tensor.
$\quad B_{ij}=-\varepsilon_{ijk}B_k$, where $\varepsilon_{ijk}$ is the Levi-Civita symbol.
The mobility tensor $\mu_{ij}=e\left(\tau/m^*\right)_{ij}$, where $e>0$, $n_i$, $\tau$, and $m^*$ are the elemental charge, carrier density, carrier relaxation rate, and effective mass, respectively. The sign $+$($-$) corresponds to the hole (electron) carriers.
The anisotropy in the electronic structure, the Fermi surface, and the relaxation rate are incorporated into the anisotropy of the mobility tensor.

The resistivity of a multi-carrier system is obtained by summing the conductivity tensors for each carrier type and taking the inverse of the total conductivity tensor:
\begin{eqnarray}
    \hat{\rho}= \left(\sum_i \hat{\sigma}_i \right)^{-1},
\end{eqnarray}
where $i$ is the index for different ellipsoids, and the Hall resistivity $\rho_H$ is determined by the off-diagonal terms (e.g., $\rho_{yx}, B\parallel z$).

It is well known that Hall resistivity exhibits a linear field dependence, $\rho_H =-B/n e$, which serves as a basis for the conventional evaluation of the AHE. However, this relation holds only for single-carrier systems. In multi-carrier systems, deviations from linearity naturally emerge due to the coexistence of carriers with different mobility tensors and densities, giving rise to a non-linear term (PAHE)~\cite{Zhang2024, Liu2024, Pi2024}.
Even in such multi-carrier systems, the derivative $\partial \rho_{H}/\partial B$ in the high-field limit remains determined solely by the total carrier density. Therefore, the non-linear component $\rho_H^{\rm NL}$ can be systematically separated from the linear component $\rho_H^{\rm L}$ as
\begin{eqnarray}
\rho_{H}&=&\rho_{H}^{\text{L}}+\rho_{H}^{\text{NL}},\quad
\rho_{H}^{\text{L}} \equiv -\frac{B}{\sum_{i}q_i n_i},
\label{eq3}
\end{eqnarray}
where $q_i$ is the charge of each carrier.
It is important to emphasize that this non-linear component $\rho_H^{\rm NL}$ is entirely distinct from the AHE and arises purely from the (anisotropic) multicarrier nature of the system.

\begin{figure}[tb]
    \centering
    \includegraphics[width=8cm]{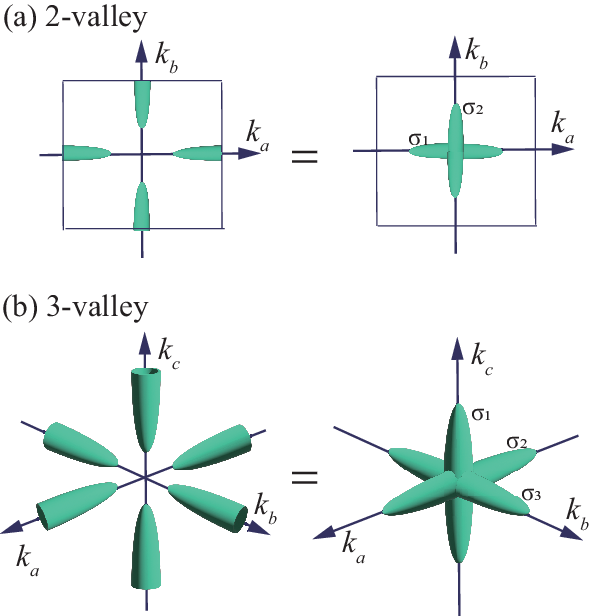}
    \caption{\label{fig_schemacic} (a) Two- and (b) Three-valley systems and the corresponding multi-ellipsoidal Fermi surfaces}
\end{figure}
\begin{figure}[tb]
    \centering
    \includegraphics[width=8cm]{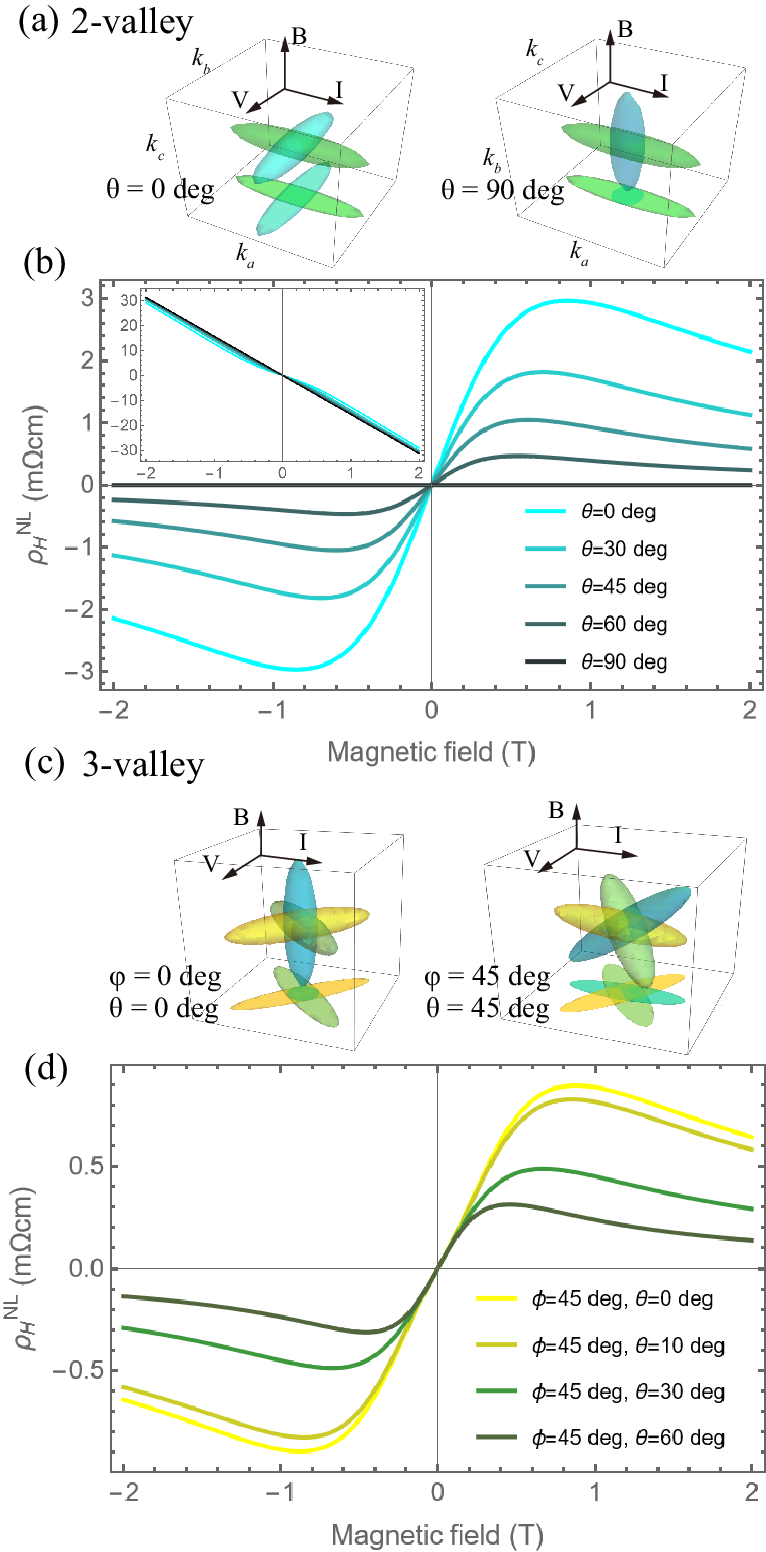}
    \caption{\label{fig_pAHE_qual_metal} (a) Non-linear component of Hall resistivity $\rho_{H}^{\text{NL}}$ (PAHE) observed for an anisotropic two-carrier system, and the total value (inset). (b) $\rho_{H}^{\text{NL}}$ observed for an anisotropic three-carrier system. $\beta=0.2,\,\mu_0=3.5\,\text{T}$}
\end{figure}

\section{Anisotropic Multivalley Semiconductors}
As a striking example of a non-linear component, we investigate semiconductors with anisotropic multivalleys, as illustrated in Fig. \ref{fig_schemacic}. In these models, the location of valleys is irrelevant to the results. The system that has valleys at the boundary of the Brillouin zone yields equivalent results to those for the system where the valleys are located at the center, as shown in Fig. \ref{fig_schemacic}. The following calculation should hold for semiconductors and metals where the effective mass approximation is valid.

\subsection{Two-valley system}
We first consider a two-valley system in which two ellipsoidal Fermi surfaces are oriented perpendicularly to each other, with the long axis of one ellipsoid aligned along the electric current shown in Fig. \ref{fig_pAHE_qual_metal}(a).
The anisotropy of the Fermi surfaces is incorporated into the mobility tensor by introducing a modulation of one component:
\begin{eqnarray}
\hat{\mu}_1=\mu_0\left( \begin {array}{ccc}
\beta  & 0  & 0\\
0  & 1  & 0\\ 
0 & 0 & 1 \end {array} \right) ,\quad \hat{\mu}_2=\mu_0\left( \begin {array}{ccc}
1  & 0  & 0\\
0  & \beta  & 0\\ 
0 & 0 & 1 \end {array} \right) ,
\end{eqnarray}
where the parameter $\beta$ expresses the anisotropy of the mobility.
Figure \ref{fig_pAHE_qual_metal}(b) shows the field dependence of non-linear Hall resistivity $\rho_{H}^{\text{NL}}$ for a two-valley semiconductor with the Fermi surface geometry depicted in Fig. \ref{fig_pAHE_qual_metal} (a). 
The Hall resistivity before subtracting the linear background is shown in the inset. In a two-valley system, the Hall resistivity exhibits a cubic dependence on the magnetic field in the intermediate-field region and asymptotically becomes linear in the high-field limit. This behavior explains the peak of $\rho_{H}^{\text{NL}}$ at low fields and its decay at high fields. 
Importantly, the two-valley system exhibits anisotropy only in the $a$-$b$ plane. 
Consequently, $\rho_{H}^{\text{NL}}$ is completely suppressed when the magnetic field is applied parallel to this plane. 
This can be experimentally achieved by rotating the samples by 90 degrees from the plane defined by the current and Hall voltage, as shown in Fig. \ref{fig_pAHE_qual_metal} (a). 

\subsection{Three-valley system}
The magnetic field and angular dependence of the Hall resistivity for a three-valley system is shown in Fig. \ref{fig_pAHE_qual_metal} (d). The mobility tensors for the three-valley system are given by
\begin{eqnarray}
\hat{\mu}_1&=&\mu_0\left( \begin {array}{ccc}
\beta  & 0  & 0\\
0  & 1 & 0\\ 
0 & 0 & 1 \end {array} \right) ,\quad \hat{\mu}_2=\mu_0\left( \begin {array}{ccc}
1  & 0  & 0\\
0  & \beta  & 0\\ 
0 & 0 & 1\end {array} \right) \\
\hat{\mu}_3&=&\mu_0\left( \begin {array}{ccc}
1  & 0  & 0\\
0  &1  & 0\\ 
0 & 0 & \beta \end {array} \right). \nonumber
\end{eqnarray}
In this configuration, the long axes of the ellipsoids are oriented perpendicular to each other. 
Under such conditions, the non-linear field dependence is not suppressed by rotating the Fermi surfaces, while it is suppressed when the projection of the ellipsoids in the current--voltage plane reduces to a single ellipsoid in the two-valley systems.
This result implies that $\rho_{H}^{\text{NL}}$ arises when the Fermi surface exhibits anisotropies in all directions within the Brillouin zone.

In conclusion, the presence of more than three anisotropic Fermi surfaces, each with a different orientation and the same carrier type, generally leads to a non-linear field dependence in the Hall resistivity.

\section{Semimetals}
We consider a semimetal with isotropic electron and hole pockets, where the mobility tensors are given by
\begin{eqnarray}
    \hat{\mu}&=&\mu_0\left( \begin {array}{ccc}
1  & 0  & 0\\
0  & 1  & 0\\ 
0 & 0 & 1 \end {array} \right) ,\quad 
\hat{\nu}=\nu_0\left( \begin {array}{ccc}
1  & 0  & 0\\
0  & 1  & 0\\ 
0 & 0 & 1 \end {array} \right),
\end{eqnarray}
for electrons and holes, respectively. Here, $\mu_0$ and $\nu_0$ are the isotropic mobilities of electrons and holes.
The total conductivity is given by
\begin{eqnarray}
    \hat{\sigma}=e\left[ n_e \left(\hat{\mu}^{-1}-\hat{B}^{-1} \right)
+n_h \left(\hat{\nu}^{-1}+\hat{B}^{-1} \right)\right],
\end{eqnarray}
where $n_e$ and $n_h$ are the electron and hole carrier densities, respectively.

\begin{figure}[tb]
\centering
\includegraphics[width=8cm]{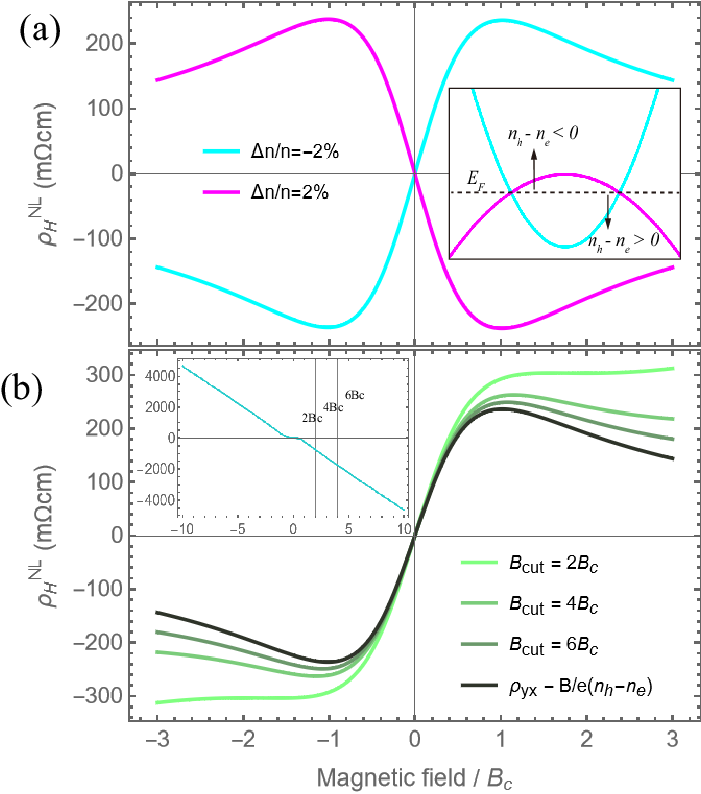}
\caption{\label{fig_pAHE_qual_semi} (a) $\rho_{H}^{\text{NL}}$ for an isotropic two-carrier semimetal with positive or negative extra carriers. The corresponding energy dispersions and Fermi energy are shown in the inset. (b) Cut-off field dependence of numerically extracted non-linear component: $\rho_{H}-B\frac{\partial\rho_{H}}{\partial B}|_{B=B_{\text{cut}}}$.}
\end{figure}

Figure~\ref{fig_pAHE_qual_semi} (a) shows the non-linear Hall resistivity $\rho_{H}^{\text{NL}}$ for this isotropic semimetal, where the parameters are fixed in the following values: {$\mu_0 =1.94\text{ T}^{-1},\quad \nu_0=1.39\text{ T}^{-1},\quad n_e+n_h\sim2.04\times 10^{18}$ cm$^{-3}$}. 
A finite deviation from perfect charge compensation due to ionized impurities is assumed, represented by a finite residual carrier imbalance $\Delta n /n = (n_h - n_e)/(n_e + n_h) \neq 0$. 
Evidently, non-linear behavior emerges when the system contains carriers with opposite signs, even in the absence of anisotropy.
The sign of $\rho_{H}^{\text{NL}}$ depends on the sign of the residual carrier, ${\rm sgn} \left( \rho_{H}^{\text{NL}}(B>0) \right) = -{\rm sgn}\left( \Delta n \right)$.

\subsection{Possible misevaluation of the non-linear component}
To accurately isolate the anomalous (or non-linear) component of the Hall effect, one needs to subtract $\rho_H^{\rm NL}$ from the total Hall resistivity $\rho_H$. Theoretically, the background linear term $\rho_H^{\rm L}$ is clearly defined by Eq. \eqref{eq3}, which can be determined in the high-field limit.
In practice, however, the background linear term is not determined at the ``ideal" high-field limit, but rather at the maximum field strength achievable in a laboratory, which may not be sufficient to correctly evaluate $\rho_H^{\rm L}$.

The ideal high-field limit depends on the degree of charge compensation in the system. 
At least, it should be much higher than the field where $\rho_{H}^{\rm NL}$ takes its maximum value.
The non-linear Hall resistivity $\rho_{H}^{\rm NL}$ reaches its maximum value at a characteristic field
\begin{eqnarray}
B_c\equiv \frac{1}{\sqrt{\mu_e\mu_h}}\frac{n}{|\Delta n|},
\end{eqnarray}
which is inversely proportional to the carrier density imbalance $\Delta n$.
When the system is nearly perfectly compensated, $B_c$ becomes extremely large, often exceeding the field range attainable in a typical laboratory.
Consequently, if the laboratory's maximum field is not sufficiently higher than $B_c$, the evaluated $\rho_H^{\rm L}$ will be systematically {misevaluated, which} can give rise to an artificial plateau, as shown in Fig. \ref{fig_pAHE_qual_semi}(b).

\begin{figure}[tb]
    \centering
    \includegraphics[width=8cm]{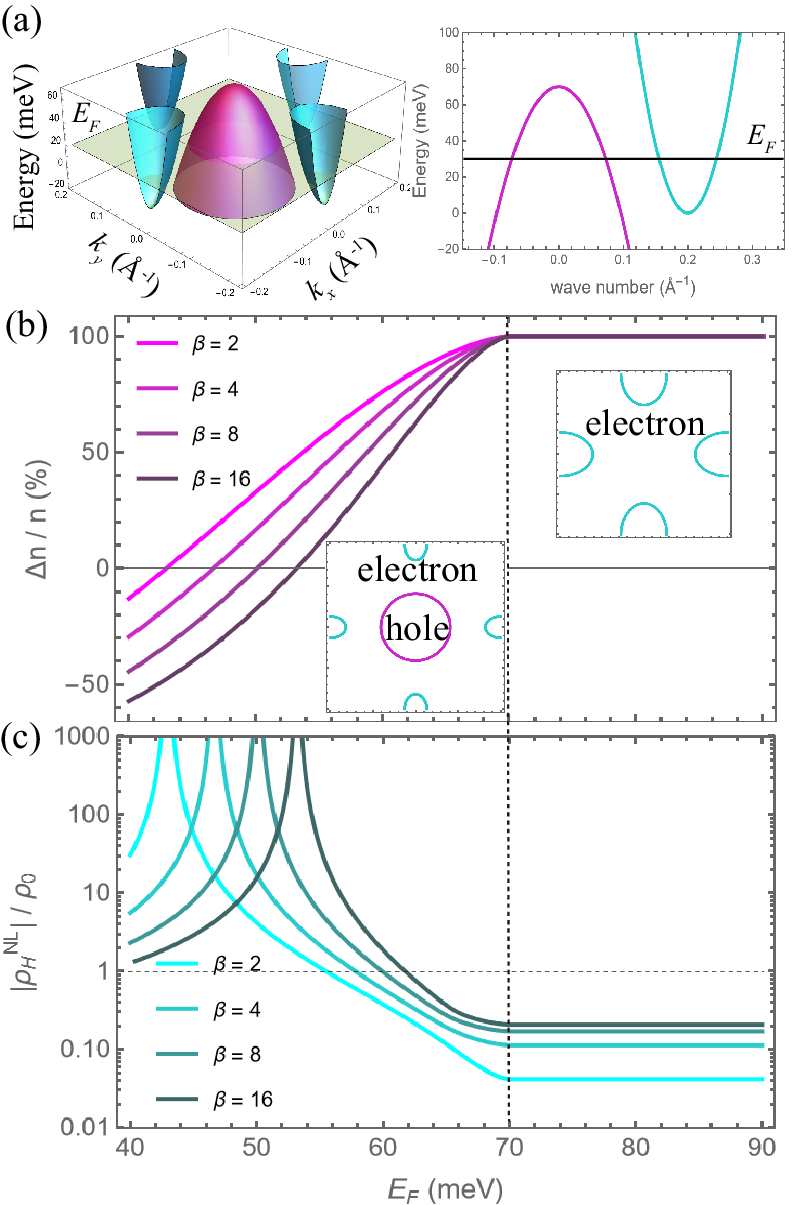}
    \caption{\label{fig_pAHE_quant_hyb} (a) Energy dispersion in a system where two anisotropic electron valleys and an isotropic hole valley coexist. The Fermi energy dependence of (b) charge polarization ($n_e-n_h/n_e+n_h$) of carrier and (c) the maximum value of $|\rho_{H}^{\text{NL}}|$ normalized by the zero-field resistivity. Anisotropy in electron carriers in the $xy$ plane is varied.}
\end{figure}

\subsection{Large non-linear term in semimetals}
The amplitude of $\rho_{H}^{\rm NL}$ in semimetals is significantly larger than that for semiconductors or metals (cf. Fig. \ref{fig_pAHE_qual_metal},\ref{fig_pAHE_qual_semi}).
This enhanced $\rho_H^{\rm NL}$ is a characteristic feature of semimetals.
To elucidate this feature more clearly, we consider an anisotropic semimetal as illustrated in Fig. \ref{fig_pAHE_quant_hyb}(a). In this model, the mobility tensors are given by 
\begin{eqnarray}
    \hat{\mu}_1&=&\mu_0\left( \begin {array}{ccc}
\beta  & 0  & 0\\
0  & 1  & 0\\ 
0 & 0 & 1 \end {array} \right) ,\quad 
\hat{\mu}_2=\mu_0\left( \begin {array}{ccc}
1  & 0  & 0\\
0  & \beta  & 0\\ 
0 & 0 & 1 \end {array} \right),\nonumber\\
\hat{\nu}&=&\nu_0\left( \begin {array}{ccc}
1  & 0  & 0\\
0  & 1  & 0\\ 
0 & 0 & 1 \end {array} \right).
\end{eqnarray}

In this model, the system behaves as a two-valley metal with only anisotropic electron pockets when $E_F$ is higher than the top of the hole band, which we set to $70$ meV.
On the other hand, for $E_F < 70$ meV, the system becomes semimetallic with coexisting electron and hole pockets.
We set $\mu_0 =1.41\text{ T}^{-1},\quad \nu_0=0.703\text{ T}^{-1}$.
In the semimetallic regime, the Fermi energy for charge neutrality, defined by $\Delta n/n = 0$, depends on the anisotropy parameter $\beta$, as shown in Fig.  \ref{fig_pAHE_quant_hyb}(b). 
As $\beta$ increases, the Fermi energy at perfect compensation $E_F$ shifts to higher energy.
Figure \ref{fig_pAHE_quant_hyb}(c) shows the maximum value of $\rho_H^{\rm NL}$ at the characteristic field $B_c$ as a function of $E_F$. The amplitude of $\rho_H^{\rm NL}$ diverges at perfect compensation ($\Delta n/n=0$) because the maximum value of $\rho_H^{\rm NL}$ is inversely proportional to $\Delta n$. 
{One can find this clearly in the isotropic semimetals in the following form:
\begin{eqnarray}
|\rho_{H}^{\rm NL}(B_c)|=\frac{ n_e+n_h}{e \sqrt{\mu  \nu } \left(n_e-n_h\right)^2 }\frac{n_e n_h (\mu +\nu )^2}{\mu\nu(n_e+n_h)^2 + (\mu n_e+\nu n_h)^2}
\end{eqnarray}}
Furthermore, even in a semimetal far from the charge neutrality, the non-linear component can exceed the zero-field resistivity, $|\rho_{H}^{\rm NL}| > \rho_0$ as shown in Fig. \ref{fig_pAHE_quant_hyb} (c).

In semimetals, the non-linear component is always included and strongly enhanced in the clean samples near the charge compensation condition.
Nevertheless, an unailablly sizeable magnetic field is needed to rigorously extract the linear components from Hall resistivity, which makes the quantitative evaluation of the non-linear component difficult in principle.

\section{ZrTe$_5$ as Anisotropic Semimetals}\label{sec_Hall_Zrte5}

\begin{figure*}[tb]
    \centering
    \includegraphics[width=17.5cm]{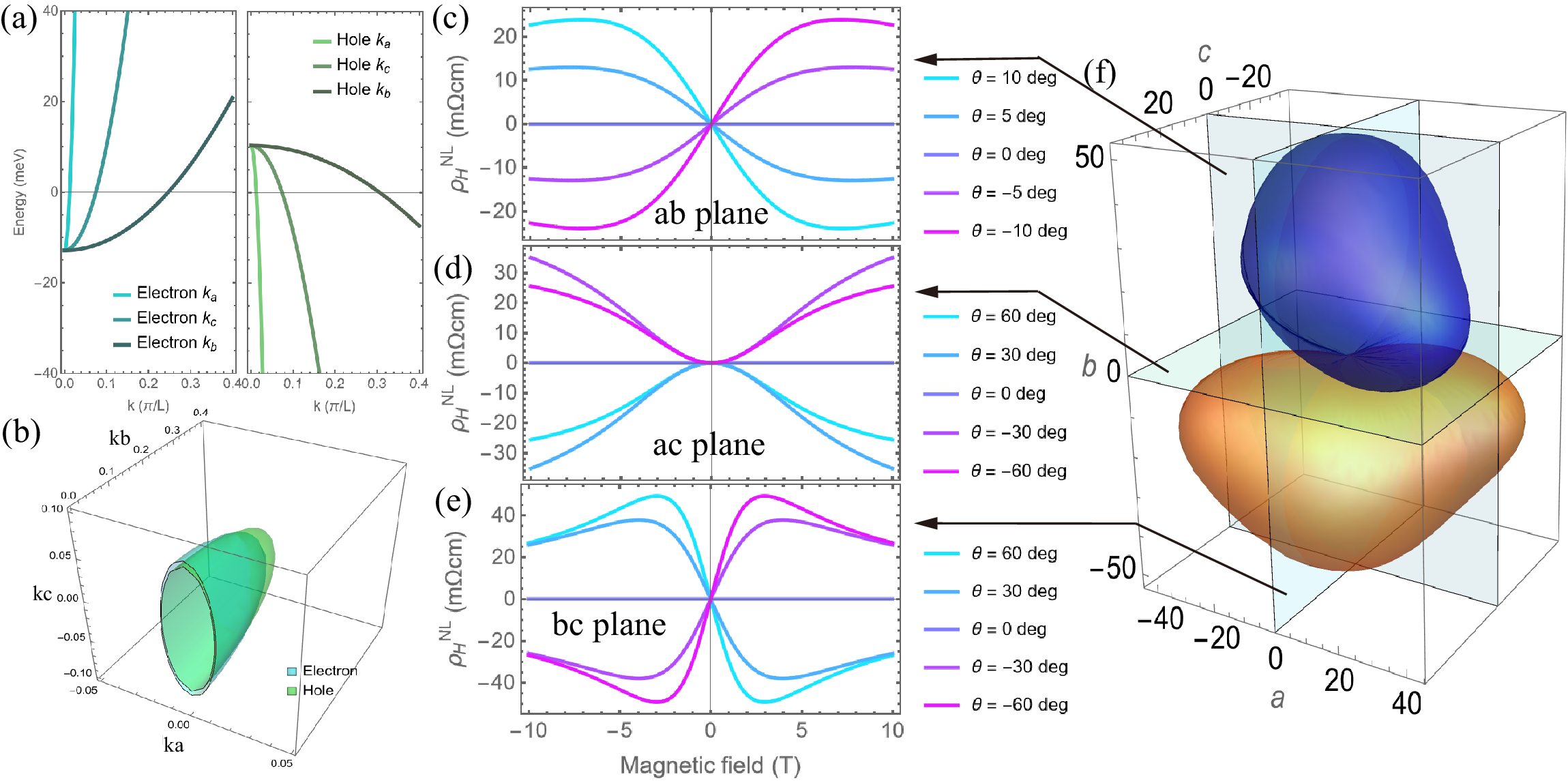}
    \caption{\label{fig_ZrTe5}(a,b) Energy dispersion model for one electron and hole designed by Kamm et al.~\cite{Kamm1985} and the Fermi surface. (c-e) Angular dependence of the non-linear components in $\rho_{ac}$ with the field rotated in (c) the ab plane ($B\parallel a$ when $\theta=0$), (d) the ac plane ($B\parallel a$ when $\theta=0$), and (e) the bc plane ($B\parallel-c$ when $\theta=0$). (f) Three-dimensional polar plot of $\rho_{ac}$, with the field fixed at $B_c$. The blue and orange areas indicate negative and positive values, respectively. The deviation from perfect compensation is assumed to be $2\,\%$. The mass anisotropy is referenced from Kamm et al., and the isotropic relaxation time is fixed to adjust $\rho_0$ along the $a$ axis to the resistivity measured by Liang et al.~\cite{Liang2018_AHE}.}
\end{figure*}

Finally, we examine the qualitative and quantitative validity of the present classical magneto-transport theory in describing the non-linear field dependence observed in ZrTe$_5$ as a practical example.
ZrTe$_5$ is considered to possess the Dirac-type dispersion with a narrow gap at the $\Gamma$ point~\cite{weng2014_spinHall_insulator,Wang2021_transportevidence_DIrac_Zrte5,Chen2015_CR,Mohelsky2023_CR}.
Its crystal structure consists of stacked two-dimensional sheets, which impart strong anisotropy to the transport properties.
ZrTe$_5$ exhibits intriguing physical properties, including the resistivity anomaly~\cite{Tritt1999_R_anomaly,Izumi_1987_R_anomaly,Kamm1985}, high thermoelectric function~\cite{wang2021_Nernst}, and quantization of Hall response~\cite{Tang2019_QHE}, drawing significant research interest. 
Non-linear field dependence or the sign change in Hall resistivity have also been reported in ZrTe$_5$~\cite{Zhang2018PRB,zhang2020,Zhang2022,Yang2020}, where the origin of this behavior in the absence of spontaneous time-reversal symmetry breaking remains unclear.

Several studies have proposed different interpretations of the carrier types in ZrTe$_5$.
Kamm {\it et al.} identified three carriers through Shubnikov-de Haas oscillations~\cite{Kamm1985}, but other experiments or theoretical models have not corroborated these findings, and the actual number of Fermi surfaces remains uncertain. 
Magneto-transport measurements suggest the presence of multiple carriers~\cite{Niu2017,Liu2021,zhuo2022,Pi2024}, including both electrons and holes, whereas some probes of Landau levels through the quantum oscillation and optical measurements suggest the presence of a single band near the Fermi level\cite{Gourgout2022_highfield,Mohelsky2023_CR}.

\begin{table}[tb]
\caption{\label{table_mobility} Components of mobility tensor in anisotropic two ellipsoids~\cite{Kamm1985}. {The unit is $10^4\,\text{cm}^2/\text{Vs}$.}}
\begin{center}
\begin{tabular}{ccc}
    \hline\hline 
    $\mu_a$ & $\mu_b$ & $\mu_c$\\ \hline
     47.1 & 4.78 & 9.89\\
    \hline\hline
\end{tabular}
\begin{tabular}{ccc}
    \hline\hline 
    $\nu_a$ & $\nu_b$ & $\nu_c$\\ \hline
    21.2 & 2.53 & 14.7\\
    \hline\hline
\end{tabular}
\end{center}
\end{table}

To model this system, we employed a semimetallic model with a single electron pocket and a single hole pocket, both centered at the $\Gamma$ point.
The mobility tensors are given by
\begin{eqnarray}
    \hat{\mu}&=&\left( \begin {array}{ccc}
\mu_a  & 0  & 0\\
0  & \mu_b  & 0\\ 
0 & 0 & \mu_c \end {array} \right) ,\quad 
\hat{\nu}=\left( \begin {array}{ccc}
\nu_a  & 0  & 0\\
0  & \nu_b  & 0\\ 
0 & 0 & \nu_c \end {array} \right),
\end{eqnarray}
where the values of $\mu_s$ and $\nu_s$ are listed in Table~\ref{table_mobility}.
The corresponding energy dispersions and Fermi surfaces are shown in Fig. \ref{fig_ZrTe5}(a) and (b).
 The ratio of mobility along the $a$, $b$, and $c$ axes is $\mu_a:\mu_b:\mu_c=9.85:1:2.07$ for the electron and $\mu_a:\mu_b:\mu_c=8.38:1:5.80$ for the hole. 
 The anisotropy of the effective mass is referenced from Kamm et al.\cite{Kamm1985}, and the carrier density and lifetime were determined by adjusting the resistivity along the $a$-axis to match experimental values~\cite{Liang2018_AHE}.
The mobility is lowest along the $b$-axis, which corresponds to the stacking direction of the two-dimensional structure. 

The current and voltage are fixed along the $a$- and $c$-axes, while the magnetic field is rotated in the $ab$-, $ac$-, and $bc$-planes.
Figures \ref{fig_ZrTe5}(c)--(e) show the field dependence of the non-linear components in the Hall resistivity for each configuration.
The corresponding three-dimensional polar plot at the characteristic field $B_c$ is shown in Fig. \ref{fig_ZrTe5}(f), where $B_c$ for anisotropic Fermi surfaces is defined as
\begin{eqnarray}
B_c&\equiv&\frac{1}{\mu_\perp }\frac{n}{|\Delta n|},\\
\mu_\perp^{-1}&\equiv& \sqrt[4]{\frac{{\bm b}\cdot\mu_e\cdot{\bm b}}{\text{det}[\mu_e]}\cdot\frac{{\bm b}\cdot\mu_h\cdot{\bm b}}{\text{det}[\mu_h]}},\nonumber
\end{eqnarray}
where $\bm b$ is a unit vector along the magnetic field.

The angular dependence in the $ab$-plane agrees with the experimental results both qualitatively and quantitatively~\cite{Liang2018_AHE}. In the $bc$- and $ac$-planes, the amplitude of the non-linear component is consistent with experiments; however, the odd component with respect to the field cannot be fully explained. In the $bc$-plane, the signs near $-60^\circ$ match the experimental value, but the absence of a sign change at $0^\circ$ cannot be explained by the anisotropy of the ellipsoidal Fermi surface.
This odd component originates from the $\sin(n\theta)$-like angular dependence of the Planar Hall effect (PHE), where $n$ is an odd number. 
A similar odd-frequency component has recently been observed in the Hall signals of a Weyl electron system~\cite{nakamura2024}.
Although multi-carrier system are known to exhibit PHE\cite{Seitz1950,Yamada2021}, the odd components are typically smaller than the even components.

A large non-linear Hall response far beyond $\rho_0$ and the anisotropy can be expected from conventional transport theory once we assume the carriers in ZrTe$_5$ consist of one electron and one hole.
The observation given by Liang et al. \cite{Liang2018_AHE} indicates the plateau-like field dependence in some direction of the field.
If the electron and hole coexist, a sufficiently high value of the magnetic field is much higher than the experimental limitation, which is the possible origin of the seemingly plateau-like field dependence (Fig. \ref{fig_pAHE_qual_semi} (b)).

\section{Conclusions}
We have demonstrated that the non-linear field dependence in the Hall effect (PAHE), which is often indistinguishable from the AHE, can be realized entirely within the framework of classical transport theory (the Drude theory).
The non-linear component $\rho_H^{\rm NL}$ originates from the anisotropy of the carrier mobility or charge compensation.
The magnitude of $\rho_H^{\rm NL}$ is inversely proportional to the residual carrier imbalance $\Delta n$, allowing it to exceed the zero-field resistivity $\rho_0$ significantly when the system is near charge neutrality.

We have highlighted the practical challenge of separating the non-linear Hall contributions $\rho_H^{\rm NL}$ from the background linear component $\rho_H^{\rm L}$.
The linear contribution $\rho_H^{\rm L}$ can only be correctly evaluated at sufficiently high fields, where $B \gg B_c \propto n/\Delta n$. 
In nearly compensated semimetals, the required magnetic field diverges as the system approaches perfect charge compensation, making it extremely difficult to isolate $\rho_H^{\rm NL}$ originating from the classical Lorentz force from other quantum effects.
Importantly, even a plateau-like structure in $\rho_H^{\rm NL}$ can emerge purely from classical mechanisms and artificial processes in numerical calculations.

As a practical example, we examined the non-linear magnetic field dependence observed in ZrTe$_5$. 
Remarkably, the anomalous contribution can be interpreted not only qualitatively but quantitatively within the classical framework for the field rotation in the $ab$-plane.
The model does not account for the odd components of the magnetic field dependence observed along other field directions.
However, this does not imply a failure of our theory. The magnitude of the classical non-linear component is comparable to the experimental values, indicating that the classical contribution cannot be neglected. Therefore, one should first evaluate the classical non-linear component before attributing any residual non-linear response to quantum effects.
The model based on parabolic dispersion employed in this study has limited applicability. 
However, it is reasonable to expect that the classical non-linear term gives a substantial contribution to the Hall effect in a wide range of materials. 

A similar example is the negative magnetoresistance.
ZrTe$_5$ is also known as one of the typical materials that show this kind of anomalous field dependence~\cite{Li2016}.
{On the other hand,} negative magnetoresistance may also arise from significant classical contributions, which have been demonstrated within the semiclassical --- not quantum --- regime~\cite{Yamada2022,Awashima_2023}. Investigating the connection between negative magnetoresistance and classical non-linear contributions to the Hall effect remains an important direction for future work.

We thank B. Fauqu\'e, M. Tokunaga, H. Sakai and K. Akiba for valuable discussions. This work was supported by JSPS KAKENHI (Grants No. 23H04862, No. 23H00268, and No. 22K18318).

\section*{References}
\bibliographystyle{iopart-num}
\bibliography{ZrTe5}

\end{document}